\documentclass[pre,aps,twocolumn,superscriptaddress,amsmath,showpacs,floatfix]{revtex4}
\usepackage{graphicx}
\begin{document}
\title{Meso-scale computer modeling of lipid-DNA complexes for gene therapy}
\author{Oded Farago}
\affiliation{Department of Biomedical Engineering, Ben Gurion University,
Be'er Sheva 84105, Israel}
\author{Niels Gr{\o}nbech-Jensen}
\affiliation{Department of Applied Science, University of California,
Davis, California 95616}
\author{Philip Pincus} 
\affiliation{Materials Research Laboratory, University of California,
Santa Barbara, California 93106}
\affiliation{Physics Department, Korea Advanced Institute of Science
and Technology (KAIST), 373-1 Kusong-dong, Yusong-gu, Taejon 305-701, 
South Korea.}
\begin{abstract}
We report on a molecular simulation method which captures the
self-assembly of cationic lipid-DNA (CL-DNA) gene delivery
complexes. Computational efficiency required for large length- and
time-scale simulations is achieved through a coarse-grained
representation of the intra-molecular details, and via inter-molecular
potentials, which effectively mimic the hydrophobic effect {\em
without}\/ explicit solvent. In addition to showing spontaneous
self-assembly of complexes, the broad utility of the model is
illustrated by demonstrating excellent agreement with X-ray
diffraction experimental data for the dependence of the spacing
between DNA chains on the concentration of CLs. At high
concentrations, the large electrostatic pressure induce the formation
of pores in the membranes through which the DNA molecules may escape
the complex. We relate this observation to the origin of recently
observed enhanced transfection efficiency of lamellar CL-DNA
complexes at high charge densities.
\end{abstract}
\maketitle

Modeling and simulation of biological systems possess significant
challenges due to the spatial complexity of such systems and by the
accompanying range of temporal scales involved. In the case of
lipid bilayers, for
instance, the width of the bilayer is in the nanometer range, whereas
the lateral dimension may extend up to several micrometers. The
corresponding time scales span an even larger range of orders of
magnitude. To address the multi-scale nature of membranes, a variety
of models have been devised which differ in the length- and time-scale
of the phenomena of interest \cite{muller_katsov_schick}. The two most
extreme approaches are: (1) atomistic computer simulations in which
the lipids and the embedding solvent are modeled explicitly in full
detail \cite{saiz_bandyopadhyay_klein}, and (2)
phenomenological models such as the effective Helfrich Hamiltonian
where the bilayer membrane is treated as a smooth continuous surface
with a small number of elastic coefficients \cite{lipowsky_sackmann}.

The gap between those two approaches can be bridged by coarse-grained
(CG) computer models in which groups of several atoms are represented
by unified particles interacting via effective potentials
\cite{nielen_lopez_srinivas_klein}. The essential benefit of such
models is that they have condensed the most detailed variations in
both time and space, while retaining the fundamental properties of
membrane self-assembly and elasticity.
The level of detail employed by different
CG models varies greatly. Some CG models are constructed to mimic
specific lipidic systems by selecting simplified representations for
water molecules and different chemical groups which constitute
the lipids, and by developing effective force-fields that reproduce
key structural features known from experiments and atomistic
simulations \cite{specific}. In coarser-grained models the
amphiphilic molecules are constructed from only two types of particles
- hydrophilic and hydrophobic - which, respectively, are attracted and
repelled from a third kind of particles representing the solvent
\cite{generic}. Such models address more general features of
self-assembly rather than specific systems.

\begin{figure}[h]
\vspace{-2.5cm}
\begin{center}
\scalebox{.40}{\centering \includegraphics{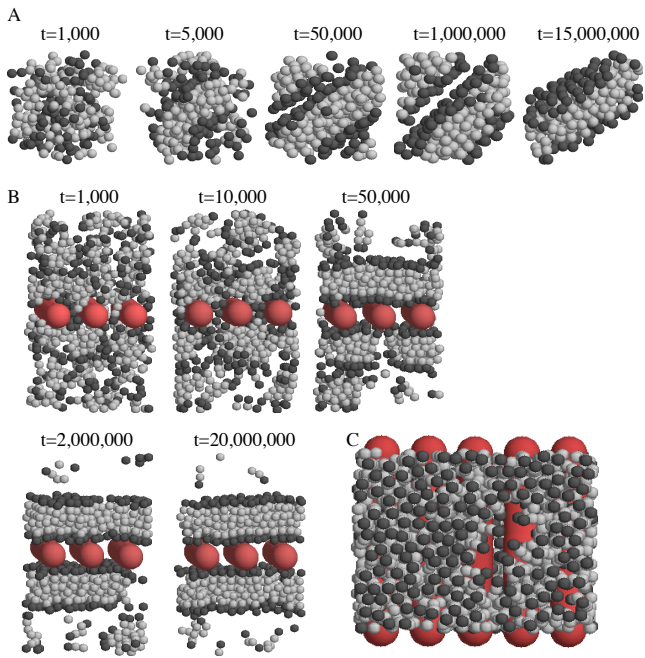}}
\end{center}
\vspace{-0.5cm}
\caption{(A) Self-assembly of a bilayer patch of 100 lipids. Lipids are 
modeled as trimers where the black and grey spheres represent
hydrophilic and hydrophobic particles, respectively (see details in
Refs.~\cite{farago}). Each Monte Carlo time step includes (on average)
an attempt to move and rotate each molecule. The membrane in the last
figure has been rotated for a better viewing of its structure. (B)
Self-assembly of a CL-DNA complex consisting of a total number of 240
lipids (of which 90 are charged). Red rods represent DNA segments. (C)
Equilibrium configuration of a complex with $\phi_c\sim0.9$ whose
membranes develop pores.}
\label{fig:selfassembly}
\end{figure}

More recently, a number of research groups have developed CG models in which
the bilayer membranes are simulated without direct representation of
an embedding solvent; this is accomplished by constructing
intermolecular force fields that mimic effects of hydration
\cite{noguchi_takasu,farago,brannigan_brown,cooke_kremer_deserno}. The
development of these ``water-free'' models is an important
accomplishment in large-scale membrane simulations, considering the fact
that the number of solvent particles in conventional CG models is
significantly larger than the number of lipids. Most of these models
employ simple Lennard-Jones (LJ) type pair-potentials
\cite{farago,brannigan_brown,cooke_kremer_deserno}, and exhibit
spontaneous self-assembly to a membrane as demonstrated in
Fig.~\ref{fig:selfassembly} (A).

The computational simplicity and efficiency of water-free CG models
make them excellent tools for testing mesoscopic scale theories of
membranes and gaining insight into the basic molecular mechanisms that
govern these systems. More importantly, these models may serve as
platforms for CG simulations of more complex systems including
bio-polymers and networks. This paper reports on such a computational
study of complexes of cationic lipid and DNA (CL-DNA) in the context
of recent experiments on molecular assemblies. The molecular
assemblies are formed spontaneously when DNA is mixed with cationic
and neutral lipids in an aqueous environment
\cite{radler_koltover_salditt_safinya1,radler_koltover_salditt_safinya2}. 
Their formation is driven by the electrostatic attraction between the
negatively charged DNA and the cationic lipid head-groups, and by the
entropic gain associated with the concurrent release of the tightly
bound counterions from the CL and the DNA
\cite{radler_koltover_salditt_safinya1,radler_koltover_salditt_safinya2,harries_may_benshaul}. 
X-ray diffraction experiments have revealed that CL-DNA complexes
exist in a variety of mesoscopic structures
\cite{safinya}, including a lamellar phase where the DNA is
intercalated in between lipid bilayers
\cite{radler_koltover_salditt_safinya1}. The DNA strands form a
one-dimensional (1D) ordered array, where the DNA interaxial spacing
$d_{\rm DNA}$ decreases with the charge density of the
membranes. Isoelectric complexes, where the charges on the DNA exactly
match those on the CL, are the most stable ones since they enable
nearly complete counterion release \cite{harries_may_benshaul}.

CL-DNA complexes have attracted much attention because of their
potential use as non-viral transfection vectors in gene therapy
\cite{safinya,felgner,clark_hersh,li_huang}. Cationic liposome 
transfer vectors exhibit low toxicity, nonimmunogenicity, and ease of
production
\cite{felgner,kamiya_tsuchiya_yamazaki_harashima}, but
their transfection efficiency (TE) remains low compared to that of
viral vectors \cite{willard,safinya2}. This has spurred an intense
research activity aimed at enhancing TE
\cite{clark_hersh,willard,safinya2}. Recognizing
that the structure of CL-DNA complexes may strongly influence their
function and TE, much of the effort in theoretical and experimental
studies has been devoted to understanding the mechanisms governing
complex formation, structure, and phase behavior
\cite{harries_may_benshaul,safinya,may_benshaul,bruinsma,harries_may_benshaul2}.

To the best of our knowledge, only a single attempt has so far been
made to study CL-DNA complexes using molecular simulations
\cite{bandyopadhyay_terek_klein}. That study employed a fully
atomistic description of both the lipids and the DNA and has,
consequently, been limited to a small system consisting of 48 lipids
(of which 20 were charged) and a short DNA segment of 10 base
pairs. The duration of the simulations was of several
nanoseconds.
The mesoscopic regime encompassing the statistics and
evolution of large molecular ensembles is, however, inaccessible to
full atomistic simulations due to the computational requirements, and
only continuum behavior of existing CL-DNA structures can be addressed
based on free energy functionals which are insensitive to the fine
details of the lipids and DNA \cite{may_benshaul}. Thus, our
understanding of many important CL-DNA complex features, such as the
process of self-assembly and membrane evolution, mesoscopic structure
and defects, can be addressed only through CG simulations.

The model presented here is based on the CG water-free membrane model,
which produces self-assembled bilayer membranes such as the one
depicted in Fig.~\ref{fig:selfassembly} (A); model details are given in
Refs.~\cite{farago}. The lipids consist of one hydrophilic
(representing the head group) and two hydrophobic (representing the
tail) spherical particles with short-range pair-interactions between
them. We set the diameter of these particles $\sigma\simeq
6.3$\AA~(see the definition of $\sigma$ in Ref.~\cite{farago}). For
this value of $\sigma$ the area per lipid in the original model
$a_{\rm lipid}\simeq 70$\AA$^2$. In the present model a fraction
$\phi_c$ of the hydrophilic head groups carry charge $+e$. DNA is
modeled as a rigid rod with a uniform axial charge density
$\lambda_{\rm DNA}=-e/1.7$\AA~and radius $R_{\rm DNA}=10$\AA. Excluded
volume interactions between rods (R) and spheres (S) are introduced
via a truncated and shifted potential of the form:
$U_{RS}/k_BT=50\left\{\left[\left(\sigma/2+R_{\rm
DNA}\right)/r\right]^{12} -1\right\}$, where $r$ is the distance
between the center of the sphere and the axis of symmetry of the rod.
The distance between nearest-neighbor rods is
restricted to $d_{\rm DNA}\geq 2R_{\rm DNA}$.

We study isoelectric complexes where the total charges of the DNA and
the CLs neutralize each other, with no added counterions. Simulations
of the quasi two-dimensional (2D) complex are conducted in a
rectangular system of size $L_x\times L_y\times L_z$, with full
periodic boundaries along the $x$ and $y$ directions \cite{niels},
and periodicity
with respect to only lipid mobility and short-range interactions in
the $z$ direction. Simulations were performed at room
temperature and with a bulk water uniform dielectric constant
$\epsilon_r=78$. The rods are
arranged in a one-dimensional array with equal spacing in the
$xy$-plane along the $y$-direction. We first examined the ability of
the complex to self-assemble spontaneously: 240 lipids, of which 90
are charged, were randomly distributed in a box of size $L_x=94$\AA,
$L_y=51$\AA, $L_z=157$\AA. Three equally-spaced DNA strands, each
carrying a total charge $-30e$, were placed at the mid-plane $z=L_z/2$
of the box. Canonical ensemble Monte Carlo (MC) simulations were used
to generate the temporal evolution of the lipids, while the positions
of the rods were fixed. Each MC step consists of (on average) an
attempt to translate (and make some minute changes in the relative
locations of the three particles with respect to each other) and
rotate each lipid. Typical evolution of the system is depicted in
Fig.~\ref{fig:selfassembly} (B). The Figure demonstrates that a
complex is formed quite rapidly, but full association of the lipids is
much slower and has not been accomplished by the end of the MC
run. The large majority of CLs (except for, typically, as few as 3-4
molecules) tend to reside in the two inner monolayers facing the DNA
array, as expected in view of the strong electrostatic attraction
between the CLs and DNA. It is possible that the final snapshot in
Fig.~\ref{fig:selfassembly} (B) is typical for an equilibrium
distribution of
free and complexed lipids, although long MC runs of pre-assembled
complexes showed no evidence for escape of lipids from the complex.

One of the parameters which is believed to strongly influence the TE
is the membrane charge density $\sigma_M=e\phi_c/a_{\rm lipid}$
\cite{safinya2}. It was
found in X-ray diffraction experiments that for isoelectric complexes
the DNA interaxial spacing, $d_{\rm DNA}$, is related to $\sigma_M$ by
\cite{radler_koltover_salditt_safinya1,safinya}
\begin{equation}
d_{\rm DNA}=\frac{\lambda_{DNA}}{\sigma_M}=
\left(\frac{a_{\rm lipid}\lambda_{\rm DNA}}{e}\right)\frac{1}{\phi_c}.
\label{eq:ddna}
\end{equation}
Measuring $d_{\rm DNA}$ and verifying this result can serve as a test
for our model's ability to mimic the meso-scale behavior of
CL-DNA complexes. We have therefore performed a set of MC
simulations of pre-assembled isoelectric complexes consisting of 300
charged lipids, an array of five equally spaced DNA molecules of
length $L_y=102$\AA~ (i.e., each carrying a total charge $-60e$), and
an additional number of neutral lipids. Simulations have been
performed using the constant surface tension ensemble $(N,\gamma,T)$,
with $\gamma=0$ \cite{brannigan_brown,remarktension}. The area of the
complex has been changed by allowing flexibility in $L_x$ (rescaling
the $x$ coordinates of particles accordingly), keeping $L_y$
fixed. Attempts to change the area of the system were made, on
average, twice every MC time unit. Two flat bilayers, each consisting
of 150 cationic and $(2N-150)$ neutral lipids, were initially set up
on both side of the DNA array with the head groups in the inner
(outer) monolayers at distance $0.5\sigma+R_{\rm DNA}$
($5.5\sigma+R_{\rm DNA}$) from the complex mid-plane. Each monolayer
consisted of $N$ lipids, with the CLs occupying the two inner
monolayers while the outer monolayers consisting of neutral lipids
only. This distribution of lipids remained unchanged over the course
of the simulations (i.e., there was no diffusion of lipids between the
monolayers on the time scale of the simulations). The fraction of
charged lipids is defined as the number ratio of charged to neutral
lipids in the inner monolayers, $\phi_c=150/N$. The pre-assembled
complexes were allowed to equilibrate for $5\times10^5$ steps prior to
data collections over, at least, $2\times10^6$ additional
steps. Membrane fluidity was verified by lateral diffusion of both
neutral and charged lipids. The former diffuse faster than the latter.

\begin{figure}[t]
\begin{center}
\scalebox{.42}{\centering \includegraphics{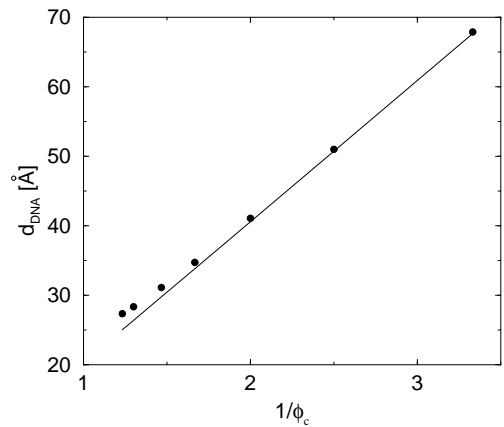}}
\end{center}
\vspace{-0.5cm}
\caption{Average DNA spacing, $d_{\rm DNA}$ as a function of the 
inverse of the fraction of charged lipids $1/\phi_c$. Circles -
numerical results (the uncertainties in the data are smaller than the
symbols); solid line - fit to Eq.~(\ref{eq:ddna}) with $a_{\rm
lipid}=69$\AA$^2$.}
\label{fig:ddna}
\end{figure}

\begin{figure}[t]
\begin{center}
\scalebox{.32}{\centering \includegraphics{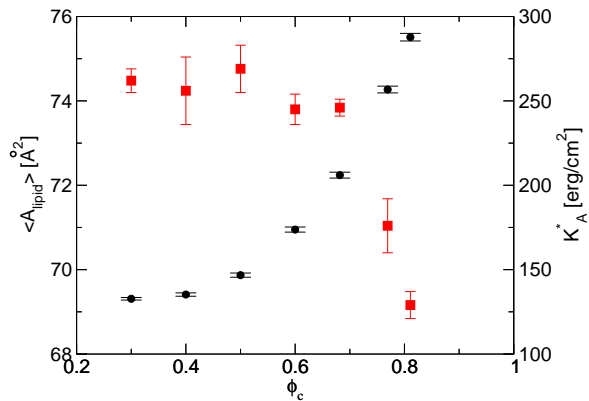}}
\end{center}
\vspace{-0.5cm}
\caption{(a) Left $y$-axis, black circle
symbols -- average area per lipid $a_{\rm lipid}$ as a function of the
fraction of charged lipids $\phi_c$. (b) Right $y$-axis, red square
symbols -- Stretching modulus $K_A^*$ of the complex as a function of
$\phi_c$.}
\label{fig:combined}
\end{figure}

The average spacing between adjacent DNA rods, $d_{\rm DNA}$, is
plotted in Fig.~\ref{fig:ddna} as a function of the inverse of the
fraction of charged lipids, $1/\phi_c$. The numerical data is in
excellent agreement with the experimental results reported in
Ref.~\cite{radler_koltover_salditt_safinya1}. The solid line is
obtained from Eq.~(\ref{eq:ddna}) with $a_{\rm lipid}=69$\AA$^2$. The
deviation from linear behavior at high charge densities arises from
the increase in the area per lipid as $\phi_c$ (see
Fig.~\ref{fig:combined}: left $y$-axis, black circle symbols). Similar
deviation is observed in the experiment
\cite{remarkagreement}. The assumption underlying
Eq.~(\ref{eq:ddna}) is that the effective interactions between the DNA
are repulsive and balanced by the elastic membrane forces. According
to Hooke's law, the elastic stress acting on a membrane is related to
$a_{\rm lipid}$ and its equilibrium value $a^0_{\rm lipid}$ by
$\tau=K_A\left(a_{\rm lipid}-a^0_{\rm lipid}\right)/a_{\rm lipid}^0$,
where $K_A$ is the 2D stretching modulus, which for lipid bilayers is
typically in the range $K_A\gtrsim 10^2\ {\rm ergs}/{\rm cm}^2$. At
high charge densities, the electrostatic stress is sufficiently large
to eliminate the membrane thermal undulations and increase $a_{\rm
lipid}$ \cite{lau_pincus}. In the present study, we find for complexes with
$\phi_c\sim0.85$ that the strain $\varepsilon\equiv\left(a_{\rm
lipid}-a^0_{\rm lipid}\right)/a_{\rm lipid}^0\sim 0.1$, which is the
typical strain lipid membranes can withstand before rapture
\cite{boal}. Membranes with higher $\phi_c$ have indeed
been found to be susceptible to pore formation, as illustrated by the
configuration in Fig.~\ref{fig:selfassembly} (C) of a complex with
$\phi_c\sim 0.9$. The loss of mechanical stability is also evident
from the rapid decrease in the complex stretching modulus $K_A^*$ for
$\phi_c\gtrsim 0.7$ (Fig.~\ref{fig:combined}: right $y$-axis, red
circle symbols), which has been extracted from the mean square
fluctuation in $a_{\rm lipid}$: $K_A^*=k_BT\,a_{\rm
lipid}^0/\left[N\left\langle\left(a_{\rm lipid}-a_{\rm
lipid}^0\right)^2\right\rangle\right]$. The larger area fluctuations
at high $\phi_c$ increase the probability of pore opening which in
turn may lead to disassociation of the complex. We suggest that this
may be the origin
of the recently observed enhanced TE of lamellar CL-DNA complexes at
high charge densities \cite{safinya2}. Transfection is viewed as a
two-stage process: (1) cellular uptake via endocytosis, and (2)
escape of the complex from the endosome, presumably through fusion of
the lipids with the endosomal membrane and release of the DNA into the
cytoplasm. TE of lamellar complexes is limited by the rate of the
second stage and, hence, increases with the decrease of mechanical
stability, i.e., with increase of charge density.

In summary, the structural-mechanical properties of CL-DNA complexes
have been studied using a coarse-grained ``water-free''
model. Despite of the highly simplified representation of lipids and
DNA, the model reliably demonstrates self-assembly of both a charge
neutral bilayer lipid membrane and a lipid membrane DNA complex. The
model reproduces accurately the measured interaxial spacing
$d_{\rm DNA}$ over a wide range of charge densities. The wide
applicability of the model should be attributed to the fact that the
meso-scale behavior of the system is dominated by non-specific
electrostatic and elastic interactions. At high charge densities we
observe that the increasing electrostatic pressure exerted on the
membranes leads to pore opening through which the DNA may be released
from the complex. The DNA release to the cytoplasm is a necessary step
for successful transfection and, thus, our results explain the
enhancement of transfection efficiency at high concentrations of
CLs. Given the consistency of agreement between our course-grained
molecular approach and observed experimental features, we suggest that
the presented model is an appropriate and promising tool for
investigating the statistics and dynamics of lipid-DNA complexes on
spatial and temporal scales relevant for biological and biomedical
applications.

We are grateful to C.~R.~Safinya for many
useful discussions. OF and PP acknowledge the support of the MRL
Program of the National Science Foundation under Award No.~DMR00-
80034 and NSF Grant No.~DMR02-037555.

\end{document}